\documentclass[aps,prd,twocolumn,10pt,nofootinbib,showkeys,superscriptaddress]{revtex4-1}
\usepackage{amsmath}
\usepackage{amssymb}
\usepackage{graphicx}
\usepackage{color}
\usepackage{hyperref}

\newcommand{\nk}{\textbf{k}}
\newcommand{\nq}{\textbf{q}}
\newcommand{\n}{\textbf{n}}
\newcommand{\nn}{\nonumber \\}
 \newcommand{\x}{\textbf{x}}

\newcommand{\bra}{\langle}
\newcommand{\ket}{\rangle}
\newcommand{\rhov}{\rho_{\textrm{vac}}}
\newcommand{\vac}{\textrm{vac}}
\newcommand{\lambdaeff}{\lambda_{\textrm{eff}}}
\newcommand{\vacio}{| 0 \rangle}
\newcommand{\barr}{\begin{eqnarray}}
\newcommand{\earr}{\end{eqnarray}}

\begin{document}

\title{Can the quantum vacuum fluctuations really solve the cosmological constant problem?}

\author{Gabriel R. Bengochea}
\email{gabriel@iafe.uba.ar} \affiliation{Instituto de Astronom\'\i
	a y F\'\i sica del Espacio (IAFE), CONICET - Universidad de Buenos Aires, (1428) Buenos Aires, Argentina}

\author{Gabriel Le\'{o}n}
\email{gleon@fcaglp.unlp.edu.ar }
\affiliation{Grupo de Astrof\'{\i}sica, Relatividad y Cosmolog\'{\i}a, Facultad
	de Ciencias Astron\'{o}micas y Geof\'{\i}sicas, Universidad Nacional de La
	Plata, Paseo del Bosque S/N 1900 La Plata, Argentina.\\
	CONICET, Godoy Cruz 2290, 1425 Ciudad Aut\'onoma de Buenos Aires, Argentina. }

\author{Elias Okon}
\email{eokon@filosoficas.unam.mx}
\affiliation{Instituto de Investigaciones Filosóficas, Universidad
	Nacional Aut\'{o}noma de M\'{e}xico, Circuito Maestro Mario de la Cueva s/n, C.U., M\'{e}xico
	D.F. 04510, M\'{e}xico.}

\author{Daniel Sudarsky}
\email{sudarsky@nucleares.unam.mx}
\affiliation{Instituto de Ciencias Nucleares, Universidad
	Nacional Aut\'{o}noma de M\'{e}xico, A.P. 70-543, M\'{e}xico
	D.F. 04510, M\'{e}xico.}
\begin{abstract}

Recently it has been argued that a correct reading of the quantum fluctuations of the vacuum could lead to a solution to the cosmological constant problem. In this work we critically examine such a proposal, finding it questionable due to conceptual and self-consistency problems, as well as issues with the actual calculations. We conclude that the proposal is inadequate as a solution to the cosmological constant problem.

\end{abstract}

\maketitle

\section{Introduction}
\label{intro}

Measurements of the brightness-redshift relation of type Ia supernovae in the 1990s led scientists to uncover the accelerated expansion of the universe \cite{Riess98, Perlmutter99, Turner98, Betoule14, Scolnic17, Abbott1811}; other observations supporting the discovery include \cite{Planck15, Eisenstein05, boss16, DES1811}. Within general relativity, the accelerated expansion can be accounted for by the inclusion of a cosmological constant, which is equivalent to the introduction of a uniformly distributed form of energy, usually referred to as ``dark energy''. Since there is no evidence for a spatiotemporal variation of such a dark component \cite{Planck15, Abbott1810, Abbott1811, Huterer17}, the cosmological constant seems to be the simplest and most favored explanation for the phenomenon\footnote{Recent analysis \cite{Riess2016, Riess2019} have uncovered certain tension among observations at different epochs that could be interpreted as a variation of the dark energy component, e.g. \cite{divalentino, zhao17, divalentino17}.}. In the next few years, substantial efforts will be made to determine if a more sophisticated dynamical scenario is required \cite{Euclid, WFIRST, LSST}.

From the theoretical point of view, considerations within quantum field theory (QFT) lead to the so-called \emph{cosmological constant problem} \cite{Weinberg89, Carroll92, JMartin12, Rugh00}. The issue amounts to a vast disagreement between the small observed value of the cosmological constant and the large theoretical prediction for the quantum vacuum energy density---which is supposed to act as a cosmological constant. The point is that the \emph{effective} cosmological constant, the one associated with the expansion rate we observe, can be naturally expected to be composed of a \emph{bare} value plus the quantum vacuum energy contribution. The problem is that the latter is calculated to be between 50 and 120 orders of magnitude larger than the value obtained from cosmological observations. As a result, in order to account for the observed value, the bare cosmological constant must be fine-tuned with extreme precision. This, in short, is the cosmological constant problem.

Recently, \cite{unruh2017} introduced a proposal for a possible solution to the problem. In such a work it is argued that, by taking seriously the \emph{non-renormalized} energy density predicted by QFT, and by assuming that it gravitates, one arrives at a constantly fluctuating and extremely inhomogeneous vacuum energy density---instead of the uniform density which is usually assumed. Such a fluctuating energy density is argued to behave differently than a cosmological constant. In particular, by treating it as an inhomogeneous stochastic field, it is supposed to lead to a spacetime that, at sufficiently small scales, oscillates between expansion and contraction. Such oscillations are however claimed to largely cancel at macroscopic scales, leaving a residual effect that, due to the weak parametric resonance of the oscillations, results in an accelerated expansion.

A more recent paper \cite{unruh2018} improved the original computational methods, allowing for the inclusion of a large number of scalar and massless fields. A higher number of fields is motivated by the fact that the Standard Model of particle physics contains several particle species, including 28 bosonic field components. In such a work it is claimed that, with the correct number of fields and an adequate cut-off, the proposed scenario leads to predictions that match observations---solving along the way the cosmological constant problem.

The aim of the present manuscript is to expose a number of serious problematic aspects of the proposal in \cite{unruh2017, unruh2018} (WZU from now on). In order to convey these, we start in section \ref{secdos} with a brief review of the standard account of the cosmological constant problem, which serves also to introduce the notation we will employ, etc. Next, in section \ref{sectres} we review the WZU model and in section \ref{seccuatro} we display what we take to be a list of severe problems with such a proposal. Finally, in section \ref{conclusions} we present our conclusions.

Regarding conventions and notation, we use a $(-,+,+,+)$ signature for the spacetime metric and units where $c=1=\hbar$.

\section{The standard account of the cosmological constant problem}
\label{secdos}

According to general relativity, the relation between spacetime and matter fields is dictated by Einstein's equations
\begin{equation}\label{EE}
G_{ab} + \lambda g_{ab} = 8 \pi G T_{ab} ,
\end{equation}
with $T_{ab}$ the stress-energy tensor of the matter fields, $G_{ab}$ the Einstein's tensor, $g_{ab}$ the metric and $\lambda$ the (bare) cosmological constant (CC).

In the context of QFT, the expectation value of the stress-energy tensor in the vacuum state is given by
\begin{equation}\label{tabvac}
T_{ab}^\vac \equiv \bra 0 | \hat T_{ab} |0 \ket = -\rhov g_{ab}
\end{equation}
with $\rhov$ a constant. The form of Eq. \eqref{tabvac} is derived from the fact that, in a flat spacetime, the vacuum is Lorentz invariant. As a consequence, the vacuum expectation of $\hat T_{ab}$ must be proportional to $\eta_{ab}$ (the Minkowski metric) as the latter is the only (0,2) tensor which is Lorentz invariant. By generalizing the previous argument to a curved spacetime, relying on the general tenants of the equivalence principle, one obtains Eq. \eqref{tabvac}, with $\rhov$ a constant due to the conservation equation $\nabla_a \bra \hat T^{ab} \ket = 0$.

Now, if we substitute $T_{ab}^\vac$ on the right hand side of Eq. \eqref{EE}, we obtain
\begin{equation}
G_{ab} + \lambda g_{ab} = 8 \pi G T_{ab}^\vac.
\end{equation}
Moving the term $8 \pi G T_{ab}^\vac$ to the left-hand side of the previous equation yields
\begin{equation}\label{EEmasvac}
G_{ab} + \lambda_{\textrm{eff}} g_{ab} = 0
\end{equation}
with
\begin{equation}
\lambdaeff \equiv \lambda + {8 \pi G} \rhov .
\label{leff}
\end{equation}
Therefore the \emph{effective} CC is the sum of the bare CC plus a contribution from the vacuum energy density.

In the context of an homogeneous and isotropic FLRW spacetime
\begin{equation}\label{frw}
ds^2 = -dt^2 + a(t)^2 (dx^2+dy^2+dz^2),
\end{equation}
with $a(t)$ the scale factor, Eq. \eqref{EEmasvac} implies $a(t) = a_0 e^{Ht}$ with $H=\sqrt{\frac{\lambdaeff}{3}}$. That is, the universe expands in an accelerated manner, with $\lambdaeff$ governing the expansion rate.

What can be said about the value of $\lambdaeff$? Theoretically speaking, going back to Eq. (\ref{leff}) we see that in order to calculate $\lambdaeff$ we need to estimate $\rhov$. To do so, one considers contributions from the zero-point energies of all fundamental quantum fields. Strictly speaking, though, the result one obtains by calculating such contributions is infinite. It is only after the introduction of an effective high-energy cut-off $\Lambda$ that one ends with a finite value for $\rhov$. Experimentally speaking, on the other hand, $\lambdaeff$ can be extracted from cosmological observations, with recent experiments \cite{Planck15} setting the value at
\begin{equation}
\lambdaeff \simeq 4.32 \times 10^{-84} (\text{GeV})^2.
\end{equation}

The problem, of course, is that the predicted value of $\rhov$ and the observed value of $\lambdaeff$ differ by between 50 and 120 orders of magnitude, depending on the assumptions of the calculation \cite{Koksma2011}. Thus, one faces the problem that, in order to match the observed and predicted values of $\lambdaeff$, one needs to fine tune the bare cosmological constant with extreme precision so as to cancel almost all, but not exactly all, the dramatically large contribution of $\rhov$. This, in a nutshell, is the CC problem \cite{Weinberg89, Carroll92, JMartin12, Rugh00}.

There have been proposals to deal with the problem by invoking protective symmetries, or similar considerations which would ensure a vanishing value for $\lambdaeff$ (see for instance \cite{Hawking1984}). The problem with said strategies, however, is that they are now invalidated by the fact that the value extracted from observations clearly does not vanish.

It should be noted that, in the above formulation of the CC problem, it was assumed that vacuum energy gravitates, i.e., that the zero-point energy encoded in $\bra \hat T_{ab} \ket$ acts as a source for the gravitational field. This is an issue on which there is an ongoing debate \cite{Weinberg89,Masso2009,Dadhich2010,rovelli1, kolb2010, JMartin12, Bernard2012, Cerdonio2015}. In fact, as noted in \cite{Weinberg, Ellis1, Ellis2, Smolin}, when addressing the problem in the semiclassical context based on unimodular gravity, vacuum fluctuations of the stress-energy tensor do not gravitate. This removes the need to contemplate the enormous discrepancy between the observed value obtained from the cosmological constant and the standard estimates from the vacuum energy. It is important to note, though, that the unimodular framework, in which a cosmological constant arises simply as an integration constant, presumably determined by initial conditions, does not, by itself, offer an explanation for the magnitude of the dark energy component inferred from cosmological observations.

\section{The WZU model}\label{unruhmodel}
\label{sectres}

In this section we present a brief summary of the WZU proposal for solving the cosmological constant problem. The model was originally presented in \cite{unruh2017} and was later revisited and improved in \cite{unruh2018}.

According to \cite{unruh2018}, the standard assumptions, which lead to a conflict between the observed value and the theoretical estimates of the CC, are the following:
\begin{enumerate}
\item The total effective cosmological constant $\lambda_{\textrm{eff}}$ is given (at the order of magnitude level) by the vacuum energy density generated by zero-point fluctuations of particle fields. In other words, vacuum fluctuations must gravitate and contribute significantly to $\lambda_{\textrm{eff}}$.
\item QFT is an effective field theory description of a more fundamental, discrete theory, which becomes important at some high energy scale $\Lambda$.
\item The vacuum expectation value of $\hat T_{ab} $ is Lorentz invariant.
\item Semiclassical gravity is valid.
\end{enumerate}
The approach proposed in \cite{unruh2017,unruh2018} to resolve the CC problem is to negate assumption 3 (see \cite{mazzi, unruh3} for a recent discussion about this point) and to replace assumption 4 with
\begin{enumerate}
	\item[4'] Semiclassical \textit{stochastic} gravity is valid.
\end{enumerate}

The starting point of the WZU analysis is the claim that, by taking seriously the quantum fluctuations of $\hat T_{ab}$ in the vacuum, one must conclude that the vacuum energy density is extremely inhomogeneous. To argue for this, they first note that the vacuum state is not an eigenstate of the local energy density operator $\hat T_{00}$, from which they argue that it must contain quantum fluctuations. To give an estimate of the size of such fluctuations, in \cite{unruh2017} they use a toy model where matter is described by a single massless real scalar field. In such a case, they find that, while $\bra \hat T_{00} \ket \propto \Lambda^4$ (recall that $\Lambda$ is an energy scale and  $\Lambda^3$  is volume$^{-1}$), the quantum fluctuations of $\hat T_{00}$, i.e.$ \bra ( \hat T_{00} - \bra \hat T_{00} \ket)^2) \ket$, are of order $2/3 \bra \hat T_{00} \ket^2 \propto \Lambda^8$.  This enormous magnitude of the quantum fluctuations associated to $\hat T_{00}$ are then argued to imply that the vacuum energy density $\rhov$ varies dramatically in space and time.

The next step in the WZU proposal is to note that the inhomogeneity of the vacuum invalidates the use of an homogeneous and isotropic FLRW spacetime. Instead they propose to use the inhomogeneous FLRW metric
\begin{equation}\label{frw2}
ds^2 = -dt^2 + a(\x,t)^2 (dx^2+dy^2+dz^2),
\end{equation}
where now the scale factor $a(\x,t)$ is a function of space and time. Moreover, they propose to model the fluctuating quantum energy density by a classical stochastic field whose stochastic properties are determined by the quantum expectation values in the vacuum. Given the metric characterized by Eq. \eqref{frw2} and a stochastic tensor $T_{ab} (\x,t)$, the dynamical equation they then consider is
\begin{equation}\label{masterunruh}
\ddot a(\x,t) + \Omega^2(\x,t) a(\x,t) = 0,
\end{equation}
where
\begin{equation}
\Omega^2(\x,t) = \frac{4 \pi G}{3} \left( T_{00} (\x,t) + \frac{1}{a^2(\x,t)} \sum_{i=1}^3 T_{ii} (\x,t) \right) .
\end{equation}

One can recognize Eq. \eqref{masterunruh} as a harmonic oscillator equation for each $\x$, with $\Omega$ playing the role of a time-dependent frequency. For the case where $\Omega(\x, t)$ is strictly periodic in time with a period $T$, the properties of the solutions of Eq. \eqref{masterunruh} have been studied in Floquet theory. Under certain conditions on $\Omega(\x, t)$, parametric resonance occurs and the general solution of Eq. \eqref{masterunruh} is
\begin{equation}\label{sol}
a(\x,t) = c_1 e^{H_\x t} P_1 (\x,t) + c_2 e^{-H_\x t} P_2 (\x,t),
\end{equation}
where $H_\x >0$, $c_1$ and $c_2$ are constants. The $P_1$ and $P_2$ are purely periodic functions of time with period $T$. They are in general functions oscillating around zero. The amplitude of the first term in Eq. \eqref{sol} increases exponentially with time while the second term decreases exponentially. Therefore, the first term will become dominant and the solution will approach a pure exponential evolution
\begin{equation}
a(\x,t) \simeq e^{H_\x t} P (\x,t)
\end{equation}
where the constant $c_1$ was absorbed into $P$ .

In the WZU proposal it is argued that, due to the stochastic nature of quantum fluctuations, the $\Omega(\x, t)$ function is not strictly periodic. However, its behavior, it is claimed, is still similar to that of a periodic function. That is, $\Omega$ is said to exhibit quasiperiodic behavior, in the sense that it is always varying around its mean value on an approximately fixed time scale. This quasiperiodic behavior of $\Omega(\x,t)$, it is claimed, should also lead to parametric resonance behavior, i.e. the solution must take the form
\begin{equation}\label{mastersol}
a(\x,t) \simeq e^{H t} P (\x,t)
\end{equation}
where $H$ is the observable global Hubble expansion rate, given by the time average of $H_{\x}(t)$,
\begin{equation}
H= \frac{1}{t} \int_0^t H_{\x} (t') dt',
\end{equation}
where $H_{\x}$ depends on the spacetime dependent frequency $\Omega(\x,t)$, but with $H$ a constant.

In addition, $P(\x,t)$ is no longer a strictly periodic function, as in Eq. \eqref{sol}, but a quasiperiodic function with the same quasiperiod as the time dependent frequency $\Omega(\x,t)$ (which is estimated to be of order $1/\Lambda$). Moreover, since $H \geq 0$ and $P$ has time average $ \overline P = 0$, taking the time average of $\dot a/a = H + \dot P/P$ yields $H = \overline{(\frac{\dot a}{a} )}$. As a result of all this, Eq. \eqref{mastersol} implies an exponentially growing scale factor, resulting in an observable distance scaling $L(t) = L(0) e^{H t}$, with macroscopic acceleration obeying $\frac{\ddot L(t)}{L(t)} = H^2$.

The rest of the WZU work is devoted to determining the specific solution $P(\x,t)$ and the value of $H^2$ associated with the relevant matter fields. If it turns out that $H^2 \sim 1$ (in Planck units), the model would not resolve the CC problem. If, on the other hand, $H^2 \sim 10^{-120}$, then the model would predict an appropriate order of magnitude for the observed acceleration.

The case of a single scalar field considered in \cite{unruh2017} leads to
\begin{equation}\label{Omega2}
\Omega^2 = \frac{8 \pi G }{3} \dot \phi^2 > 0,
\end{equation}
where $\phi$ is a classical stochastic field whose statistical properties are determined by the quantum fluctuations of the vacuum. It is crucial for the WZU model that $\Omega^2$ is positive, otherwise the observed expansion would not be correctly described by the model. In \cite{unruh2018}, a more realistic model is developed and the numerical calculations are improved. In particular, they find that, for a universe with 28 bosonic fields (as in the Standard Model of particle physics), and with a high-energy cut-off $\Lambda$ 40 times higher than the Planck energy, they obtain $H \sim 10^{-60}$, which is comparable to the observed value. Thus, according to the authors of the WZU model, by taking seriously the quantum fluctuations of the vacuum, we might be able to solve the cosmological constant problem.

In order to get a feel for the procedures employed by WZU to arrive at these conclusions, we start by considering the case of a single massless scalar field. For $\hat \phi$ a quantum massless scalar field in Minkowski spacetime, one has
\barr\label{phimink}
\hat \phi (\x,t) &=&\int \frac{d^3 k}{(2 \pi)^{3/2}} \frac{1}{\sqrt{2 \omega_k}} \bigg[ \hat a_{\nk} e^{ -i(\omega_k t - \nk \cdot \x)}+\nn
&+& \hat a_{\nk}^{\dagger} e^{ +i(\omega_k t - \nk \cdot \x) } \bigg],
\earr
where $\omega_k = |\nk|$. The vacuum state $|0\ket$ is then defined by $\hat a_{\nk} |0 \ket = 0 $ for all $\nk$. According to \cite{unruh2018}, since Eq. \eqref{masterunruh} contains no spatial derivatives, one can consider a fixed point in space and focus only on the time evolution of $a$. Therefore, for a fixed $\x_0$ one has
\begin{equation}\label{phit}
\hat \phi (\x_0,t) = \int \frac{d^3 k}{(2 \pi)^{3/2}} \frac{1}{\sqrt{2 \omega_k}} \left( \hat b_{\nk} e^{ -i\omega_k t} + \hat b_{\nk}^{\dagger} e^{ +i\omega_k t } \right),
\end{equation}
where $\hat b_{\nk} \equiv e^{i \nk \cdot \x_0} \hat a_{\nk}$. As is done in \cite{unruh2018}, we omit from now on the label $\x_0$ and rewrite Eq. \eqref{phit} as
\begin{equation}
\hat \phi (t) = \int \frac{d^3 k}{(2 \pi)^{3/2}} \frac{1}{\sqrt{2 \omega_k}} \left( \hat x_{\nk} \cos(\omega_k t) + \frac{1}{\omega_k} \hat p_{\nk} \sin(\omega_k t) \right),
\end{equation}
with
\begin{equation}
\hat x_{\nk} \equiv \sqrt{\frac{1}{2 \omega_k}} \left( \hat b_{\nk}^\dagger + \hat b_{\nk} \right),
\end{equation}
\begin{equation}
\hat p_{\nk} \equiv i \sqrt{\frac{\omega_k}{2}} \left( \hat b_{\nk}^\dagger - \hat b_{\nk} \right).
\end{equation}	
	
Next, the Wigner-Weyl description of quantum mechanics is adopted. In this framework, any state can be represented by a quasi-distribution function, called Wigner's function. For the vacuum state characterizing the field $\hat \phi(t)$, one can construct the Wigner function $W(\{x_{\nk}\}, \{p_{\nk}\},t)$, where $\{x_{\nk}\}$ denotes the set $\{x_{\nk_1},x_{\nk_2}, \dots \} $ with all $\nk$. The resulting Wigner function is a product of Gaussians
\begin{equation}\label{wignervac}
W(\{x_{\nk}\}, \{p_{\nk}\},t) = \frac{1}{\pi} \prod_{\nk} e^{-p_{\nk}^2 - x_{\nk}^2}.
\end{equation} 	
In the Wigner representation, the field $\dot{\hat{\phi}}^2 (t)$ can be expressed as a function $\dot \phi^2(\{x_{\nk}\}, \{p_{\nk}\},t)$. In particular, for $\Omega^2$ (see Eq. \eqref{Omega2}), one has
\begin{eqnarray}\label{Omega2wigner}
&&\Omega^2(\{x_{\nk}\}, \{p_{\nk}\},t) = \nn
&& \frac{8 \pi G}{3} \int \int \frac{d^3k d^3q}{(2\pi)^3} x_\nk x_\nq \omega_k \omega_q \sin \omega_k t \sin \omega_qt +\nn
&+& p_\nk p_\nq \cos \omega_k t \cos \omega_qt - 2 x_\nk p_\nq \omega_k \sin \omega_k t \cos \omega_qt .
\end{eqnarray}

The next crucial step is to assume that Eq. \eqref{masterunruh} has an equivalent equation for the quantum operators, i.e. $\ddot{\hat{ a}} (t) + \hat{\Omega}^2(t) \hat{ a} (t) = 0$, which in the Wigner representation takes the form
\begin{eqnarray}\label{masterunruhwigner}
& & \ddot a + \Omega^2 a+ \frac{i}{2} \sum_k \left( \frac{\partial \Omega^2}{\partial x_{\nk}} \frac{\partial a}{\partial p_{\nk}} - \frac{\partial \Omega^2}{\partial p_{\nk}} \frac{\partial a}{\partial x_{\nk}} \right) \nn
&-& \frac{1}{8} \sum_{{\nk},{\nk'}} \bigg( \frac{\partial^2 \Omega^2}{\partial x_{\nk} \partial x_{\nk'}} \frac{\partial^2 a}{\partial p_{\nk} \partial p_{\nk'}} + \frac{\partial^2 \Omega^2}{\partial p_{\nk} \partial p_{\nk'}} \frac{\partial^2 a}{\partial x_{\nk} \partial x_{\nk'}} \nn
&-& 2 \frac{\partial^2 \Omega^2}{\partial x_{\nk} \partial p_{\nk'}} \frac{\partial^2 a}{\partial p_{\nk} \partial x_{\nk'}} \bigg) =0
\end{eqnarray}
with $a$ and $\Omega^2$ the corresponding Wigner transforms [i.e. $a(\{x_{\nk}\}, \{p_{\nk}\},t) $ and $\Omega^2(\{x_{\nk}\}, \{p_{\nk}\},t) $].

The numerical analysis begins by discretising $\nk$ and randomly sampling $\{ x_\nk \} $ and $\{ p_\nk \}$ with the distribution of Eq. \eqref{wignervac}. The discretization is done by considering a cube of width $L$ in physical space, and restricting the allowed modes of the field to be harmonics modes of the box. The frequency of such modes is $\omega = 2\pi |\n| / L$, with $\n = (n_x,n_y,n_z)$ a set of integers. The cutoff frequency $\Lambda$ induces a cutoff on $\n$ given by $n_{\textrm{max}} = L \Lambda/2\pi$. The cutoff in momentum space is applied as a sphere of radius $\Lambda$ by choosing modes with $|\n| < n_{\textrm{max}} $. Therefore, the sets $\{x_{\nk}\}, \{p_{\nk}\}$ are now labeled as $\{x_{\n}\}, \{p_{\n}\}$, and they each contains one random number for every value of $\n$, such that $|\n| < n_{\textrm{max}} $. After randomly sampling $\{x_{\n}\}, \{p_{\n}\}$, the Wigner transform of $\Omega^2$ can be obtained from the discrete equivalent of Eq. \eqref{Omega2wigner}, which is
\begin{equation}
\Omega^2(\{x_{\n}\}, \{p_{\n}\},t) = \left[ \sum_{\n} \sqrt{n} ( x_{\n} \sin(nt) - p_\n \cos(nt) ) \right]^2 .
\end{equation}

With the expression of $\Omega^2(\{x_{\n}\}, \{p_{\n}\},t)$ obtained in the aforementioned manner, the authors of \cite{unruh2018} solve Eq. \eqref{masterunruhwigner} for $a(\{x_{\n}\}, \{p_{\n}\},t)$. The full procedure is repeated for $N$ different sets of random numbers $\{ x_\n \} $ and $\{ p_\n \}$. Subsequently, they average the $N$ solutions $a(\{x_{\n}\}, \{p_{\n}\},t)$ and identify the classical value $a_o(t )$ with such an average. Additionally, they propose to identify the observed value of $H$ with the average $\overline{\dot a/a}$ obtained from the $N$ computed solutions $a(\{x_{\n}\}, \{p_{\n}\},t)$.

According to \cite{unruh2018}, if both $N$ and $L$ increase, then the average obtained from this method should converge to the quantum expectation value of the operator $\hat a$, which can be computed analytically from the Wigner distribution and the Wigner transform of $a$. That quantum average is then identified with the classical value $a_o$. It is also noted that increasing the number of fields $n_f$, results in a total $\Omega^2$ that is simply the sum of each individual $\Omega^2_j$, i.e. $\Omega^2 = \sum_{j}^{n_f} \Omega^2_j$.

The employed method is argued to imply that the average $\overline{\dot a/a}$ over the $N$ solutions $a(\{x_{\n}\}, \{p_{\n}\},t)$ converges to the quantum expectation value of an operator $ \widehat{\dot a/a}$ in the appropriate limit of large $N$ (and $L$). The justification for such an implication is that the quantum expectation value of any operator can be calculated from the Wigner description of quantum mechanics. In particular, for the operator $\widehat{\dot a/a}$ one has
 \begin{equation}\label{wtdotaa}
 \bra \widehat{\dot a/a} \ket = \int \prod_{\nk} dx_\nk dp_\nk ( \dot a/a) [\{x_{\nk}\}, \{p_{\nk}\},] W[\{x_{\nk}\}, \{p_{\nk}\},t].
 \end{equation}
The outlined procedure yields the main plots and results of Ref. \cite{unruh2018}.

\section{Problematic aspects of the WZU model}\label{issuesunruh}
\label{seccuatro}

In this section we expose what we take to be the main problematic aspects of the WZU proposal. We start by scrutinizing the claim that the vacuum is highly inhomogeneous, then we explore the way the allegedly inhomogeneous vacuum is handled via stochastic semiclassical gravity and we end by dissecting some aspects of the calculations underlying the WZU proposal.

\subsection{Quantum fluctuations and inhomogeneities} \label{secQF}

As we mentioned in the previous section, the starting point of the WZU account is the claim that the quantum fluctuations of the vacuum imply a highly inhomogeneous vacuum energy density--- an idea which is crucial for their whole construction. In this subsection, however, we show such a statement to be deeply problematic. In order to do so, it is useful to be precise regarding the rules and assumptions at play. In particular, we find it convenient to begin by explicitly stating the postulates of quantum mechanics (see for instance, \cite{Dirac1930, newmann, albert, bassi2003}), which can be summarized as follows:
\begin{enumerate}
\item[(i)] To every \emph{quantum system} corresponds a Hilbert space.
\item[(ii)] The complete \textit{physical state} of the system is represented at all times by a unit vector in the Hilbert space.
\item[(iii)] The \textit{physical properties} of the system are represented by Hermitian operators.
\item[(iv)] The time evolution of the system is governed by a linear, unitary and deterministic equation (e.g., the Schr\"odinger equation).
\item[(v)] Upon a \emph{measurement}, the Born rule provides a list of possible results and their probabilities.
\item[(vi)] After a \emph{measurement}, the state of the system instantaneously jumps to the eigenstate of the measured property with the eigenvalue corresponding to the measured value.
\end{enumerate}

One of our concerns with the WZU model, as we will see, is that it seems to inadvertently conflict with these postulates. Let us explore the issue in detail.

We start by making a fairly obvious observation. The WZU model takes the initial state of the universe to be the vacuum $\vacio$. Such a description, according to postulate (ii), is assumed to be complete. Now, it is straightforward to check that the vacuum is completely homogeneous and isotropic, i.e., that such a state is annihilated by the generators of spatial translations and rotations. Additionally, it is easy to confirm that the unitary evolution applied to $\vacio$ maintains at all times the original homogeneity and isotropy of the system. Therefore, the physical state of a system, fully characterized by the vacuum state, is perfectly homogeneous and isotropic at all times.

The situation, of course, would radically change if one could somehow rely on postulates (v) and (vi). That is, if one could argue that some kind of \emph{measurement} was performed on the system, upon which the vacuum changed to a new state, say $| \Omega \ket $, that need not be homogeneous and isotropic $ \vacio$. The inhomogeneous and anisotropic new state $| \Omega \ket$ could then be used to characterize an inhomogeneous spacetime and matter fields after the time of measurement. The problem, of course, is that in order to employ postulates (v) and (vi), it is necessary to introduce some sort of external observer, which seems impossible in cosmological context at play. It is clear, then, that in order to obtain an inhomogeneous state $| \Omega \ket$ from the symmetric vacuum, without invoking observers or measuring apparatuses, one must depart from the standard interpretation of quantum mechanics described above.

In spite of all this, the WZU position is that, even in the complete absence of measurements or observers, the quantum fluctuations of the vacuum imply it being inhomogeneous. To ague for this they begin by noting that the vacuum state, although an eigenstate of the global Hamiltonian $\hat H = \int d^3x \hat T_{00}$, is not an eigenstate of the local energy density operator $\hat T_{00}$. From there they claim that the inhomogeneities arise from quantum fluctuations encoded in the covariance of the energy density operator, which is defined as
\begin{eqnarray}\label{covTab}
& & \textrm{Cov}(\hat T_{00} (x), \hat T_{00}(y) ) =\bra 0| \{ ( \hat T_{00} (x) - \bra 0| \hat T_{00} (x) | 0 \ket ) \nn&\times& ( \hat T_{00} (y) - \bra 0| \hat T_{00} (y) | 0 \ket ) \} |0 \ket ,
\end{eqnarray}
where the curly brackets $ \{ \} $ indicate symmetrization. If $x$ and $y$ are equal, Eq. \eqref{covTab} yields the variance of $\hat T_{00}$, which is of the same order of magnitude as $\bra \hat T_{00} \ket^2, $ i.e. $(\bra \hat T^2_{00} \ket - \bra \hat T_{00} \ket^2 ) \sim \bra \hat T_{00} \ket^2 \sim \Lambda^8$. Additionally, from the fact that when the spatial distance between $x$ and $y$ increases, then $\textrm{Cov}(\hat T_{00} (x), \hat T_{00}(y) ) \to 0 $, it is concluded that the fluctuation at distant $x$ and $y$ are independent. All these results are taken to indicate that the quantum vacuum is extremely inhomogeneous.

A general problem with all this is that it is not clear how one could arrive at the conclusion that the vacuum energy density is inhomogeneous by inspecting quantities, such as $\bra 0 | \hat T_{00} ( x) | 0 \ket $ and $\bra 0 | \hat T_{00}^{2} ( x) | 0 \ket $, which can be formally shown to be independent of $x$, that is, exactly homogeneous and isotropic. Still, let us explore the WZU argument in more detail.

In order to argue that the vacuum is inhomogeneous, WZU holds that the fluctuation at distant points are independent. It is easy to see, however, that this cannot be correct. The total energy is given by the integral of the energy density over all points, but if the fluctuations at different points were uncorrelated, the integral would be equivalent to a random walk---which generically differs from zero. This, of course, is incompatible with the fact that the vacuum is an eigenstate of the Hamiltonian with eigenvalue 0. To see this in a simpler system, consider a pair of spin-$\frac{1}{2}$ particles in a singlet state. In analogy with the WZU argument, one might claim that, since such a state is not an eigenstate of the spin of each of the particles, then such quantities would have fluctuations. If so, as each spin has magnitude $\frac{1}{2}$ with a randomly fluctuating direction, the total angular momentum would range from 0 to 1. This, of course, would be incompatible with the fact that the total spin of the singlet is exactly zero.

What is wrong, then, with the argument by which one starts from "taking into account that when the spatial distance between $x$ and $y$ increases, $\textrm{Cov}(\hat T_{00} (x), \hat T_{00}(y) ) \to 0 $" and then concludes that the fluctuation at distant $x$ and $y$ are independent? The problem is that the covariance goes to zero not because the correlations disappear but because they get ``diluted'', as more and more points are involved in the correlation as the distance between points grows.

 On the other hand, when the spatial distance between $x$ and $y$ decreases, the quantity Cov$(\hat T_{00} (x), \hat T_{00}(y) ) $ increases (specifically, in Ref. \cite{unruh2017} is shown that it increases towards $2/3 \bra \hat T_{00}  \ket^2$ when the spatial distance goes to zero). In the WZU model, that fact is taken to support the argument that the energy density, which is associated with the operator $\hat T_{00}$, is extremely inhomogeneous. To show that this cannot be the case, let us explore the issue more generally. In quantum theory, a two-point function $\bra \hat A(x) \hat B(y) \ket $ is a quantum correlation between the values of two operators, $\hat A$ and $\hat B$, associated with events $x$ and $y$ respectively. Given this, one may wonder if a non-vanishing value of $\bra \hat A(x) \hat B(y) \ket $ could imply some sort of inhomogeneity, that is, if it could signal that something about the state of the world is different in $x$ and $y$. In order to clarify the issue, let us again examine the question in the much simpler EPR-B scenario.

Consider the decay of a spin-0 particle at the origin, taking place along the $z$ axis. The joint state of the two spin-$\frac{1}{2}$ particles is a singlet state $|\Psi \ket$, which is invariant under rotations around the axis of decay. Now, let us consider vectors $\vec n_1$ and $\vec n_2$ perpendicular to the $z$ axis, and construct the operators $\hat A$, the spin of particle 1 along direction $\vec n_1$, and $\hat B$, the spin of particle 2 along direction $\vec n_2$. It is easy to see that there is a non-vanishing quantum correlation between $\hat A$ and $\hat B$. In fact, $\bra \Psi | \hat A \hat B | \Psi \ket $ is proportional to $\vec n_1 \cdot \vec n_2 = \cos (\theta)$ where $\theta$ is the angle between the two orientations. The question we are interested in is if we can take this nontrivial two-point correlation as an indication that the symmetry under rotation around the $z$ axis has been broken.

One might get the impression that this is the case by assuming that the correlation somehow means that particle 1 now has a spin along the $\vec n_1$ axis (even if the sign is still unknown to us) and that particle 2 now has a spin along the $\vec n_2$ axis. However, what the correlation in fact indicates is that if and when we decide to measure those spins, the results over a long series of repeated experiments would lead to statistical correlations between the two sets of results that would go as $\cos (\theta)$. Moreover, in the absence of a measurement, the answer to the question above is negative. That is, in the absence of a measurement, the state remains $|\Psi \ket$, a fully rotationally invariant state, in which neither particle 1 nor particle 2 has a definite spin in any direction. It goes to the core of quantum mechanics that two things can be correlated, despite not having definite values.

Let us now focus on so-called quantum fluctuations, which are associated with two-point functions, in the particular case when the two operators are equal and are evaluated at the same point, i.e., $\bra \hat \phi^2 (x) \ket $. It is well known that if $\hat \phi$ is a quantum field, then $\hat \phi(x) $ is not defined as an operator on each point $x$ of the spacetime. In fact, $\hat \phi$ is formally well-defined only as a distribution on spacetime, but the product of two distributions at the same spacetime point, e.g., $\hat \phi^2(x)$, is intrinsically ill-defined mathematically. As a consequence $\bra \hat \phi^2 (x)\ket $ by itself is divergent. Nonetheless, there are physical observables in QFT that depend on the expectation values of $\hat \phi^2 (x)$. In particular, for $\bra \hat T_{ab} (x) \ket$, which depends quadratically on the fields, one can construct a well-defined renormalization procedure to obtain meaningful results. It is only in this context where one can assign some kind of meaning to $\bra \hat \phi^2 (x)\ket $. At any rate, the relevant conceptual aspects of the issue at hand can be explored in a more elementary scenario, namely the ground state of the simple harmonic oscillator.

The point we want to make is that an identification between quantum fluctuations and actual, physical inhomogeneities, is questionable (or at least incomplete). The problem is that the quantum fluctuations cannot be taken to represent \emph{physical fluctuations}, as they are only a measure of the \emph{width} of the quantum state in question. To see this, consider the ground state of a 1D simple harmonic oscillator, which clearly has uncertainty in position. The crucial point, however, is that such an uncertainty does not imply that the ground state is not symmetric under a reflection along the origin; instead, the uncertainty is only a measure of the spread of the results of several position measurements, performed on an ensemble of identically prepared systems. As a result, in order to break the reflection symmetry of a single harmonic oscillator, an actual measurement of position has to be performed. In other words, the quantum fluctuations or uncertainties do not, by themselves, indicate that some aspect of the physical system is undergoing random or stochastic motion, and as far as a quantum state of the system is taken to describe it completely, the symmetries of the quantum state must be taken as also completely characterizing the system to which such a state is associated.

Similarly, the fluctuations or uncertainties in the vacuum considered by WZU do not, in any way, constitute a departure from homogeneity or isotropy. Without an actual, physical process, beyond that imposed by the unitary dynamics (which clearly does not break such symmetries), no deviation from the initially homogeneous state can occur. And since, as we discussed above, no measurements can happen in this setting, clearly there is something missing in the WZU account.

We conclude that any stochasticity attributed to the vacuum necessarily requires the identification of an observer and/or a measurement device external to the system. Since it seems impossible to identify such entities in the cosmological setting, the inhomogeneities considered by WZU remain obscure. It is often argued that decoherence---i.e., the inevitable interaction of a system with its environment---is able to explain the quantum-to-classical transition. If so, one might argue that decoherence is responsible for the surge of inhomogeneities in the vacuum. The problem with all this is that decoherence by itself is in fact incapable of explaining this transition, \cite{adler, schlosshauer, Bub1997, Bacciagaluppi, okon15}. Decoherence operates within the framework of standard, linear, unitary quantum mechanics. Therefore, it cannot destroy by itself superpositions or symmetries.

To see this, we note that the argument for the claim that decoherence can explain the quantum-to-classical transition is that, for all practical purposes, reduced density matrices of systems in interaction with an environment behave as mixtures. The problem is that those reduced density matrices behave as mixtures \emph{only if} one assumes that, upon measurement, systems behave according to the Born rule and the collapse postulate, i.e., according to postulates (v) and (vi). Consequently, decoherence alone, without any external input that might be recognized as a measurement, cannot provide a justification for a stochastic description of the system under examination. At any rate, the universe, by definition, is an isolated system. Therefore, no clear candidates for environmental degrees of freedom to be traced out seem to be available, \cite{bassi2013}.

Of course, one might want to go beyond the standard postulates of quantum theory (and indeed one needs to do so if one wants to work within a framework that is not plagued with conceptual difficulties). But in that case, one has to clearly specify what alternative approach to quantum theory one is using. Otherwise, one is simply utilizing a collection of mutually incompatible premises, choosing one at a time according to what one needs to achieve at the corresponding stage.

\subsection{Semiclassical gravity and stochasticity} \label{secSSCG}

The WZU model, relies on semiclassical ideas for the treatment of gravity (in the sense of describing spacetime geometry in terms of a classical metric, while characterizing matter fields in term of quantum theory). Traditional semiclassical gravity (SCG) is based on Einstein's semiclassical equations \cite{rosenfeld1963}
\begin{equation}\label{scg}
G_{ab} = 8 \pi G \bra \psi | \hat T_{ab} | \psi \ket.
\end{equation}
A natural reading of the SCG approach assumes that spacetime is quantum mechanical at the fundamental level, but considers that when a metric characterization is meaningful, one is already well within the classical realm as far as the gravitational degrees of freedom are concerned. In other words, SCG must be seen as an effective theory and not as a fundamental one.

There are, however, some known situations in which Eq. \eqref{scg} fails even as an effective theory, such as when the quantum uncertainties of $\hat T_{ab}$ are large compared to its expectation value. According to the WZU model, this is the case in the cosmological setting. Consequently, in such a model, SCG is replaced by \textit{stochastic} semiclassical gravity (SSCG). That is the main idea behind the assumption 4' described in Sect. \ref{sectres}.

The motivation behind the SSCG framework is to take into account the effects on spacetime of the quantum fluctuations of the stress-energy tensor. One of the most well-known approaches to SSCG is developed in Refs. \cite{verdaguer1998a,verdaguer1998b,verdaguer1999a,verdaguer1999b,verdaguer2007,verdaguer2008}. The proposal is to consider the spacetime metric as an open system that interacts gravitationally with the quantum matter fields, the latter constituting the environment. As a consequence, the system will exhibit stochastic dynamics with fluctuations due to the noise induced by the environment. For simplicity, a perturbative analysis is usually considered. To the lowest order, SSCG is thus characterized by the modified Einstein's semiclassical equations
\begin{equation}\label{sscg}
G_{ab} [g+h] = 8 \pi G \bra \hat T_{ab} [g+h] \ket_R + 8\pi G \xi_{ab} (g),
\end{equation}
where $g$ is a metric that results from solving the standard Einstein's semiclassical equations, $h$ is a linear perturbation and $\bra \hat T_{ab} \ket_R $ refers to the renormalized stress-energy tensor. The field $\xi_{ab} [g]$ is a Gaussian stochastic classical noise; its statistical properties are inherited from the quantum fluctuations of the stress-energy tensor and are taken to be
\begin{subequations}\label{sscgeq}
\begin{equation}
\bra \xi_{ab} (x) \ket_s = 0,
\end{equation}
\begin{equation}
 \qquad N_{abcd} (x,y) \equiv \bra \xi_{ab} (x) \xi_{cd} (y) \ket_s = \bra \{ \hat t_{ab} (x) \hat t_{cd} (y) \} \ket [g]
\end{equation}
\end{subequations}
where $\hat t_{ab} (x) \equiv \hat T_{ab} (x) - \bra \hat T_{ab} (x) \ket$ and higher order cumulants are set to zero. It is important to point out that two notations are being introduced: $\bra \ldots \ket_s$ and $\bra \ldots \ket$. The notation $\bra \ldots \ket_s$ refers to an average associated to a classical stochastic process. That is, an average over a suitable ensemble of ``possible realizations'', with the understanding that each individual experiment corresponds to a single unique realization (thus, in the cosmological setting at hand, our universe would correspond to a single realization). The notation $\bra \ldots \ket$ refers to the quantum expectation value of an operator. As can be observed from Eqs. \eqref{sscg} and \eqref{sscgeq}, the stress-energy quantum fluctuations induce a back-reaction effect on the spacetime geometry. Specifically, the term $\xi_{ab}$ induces a perturbative correction to semiclassical gravity. Thus, it is assumed that the gravitational field is described by $g_{ab} + h_{ab}$, with $h_{ab}$ a linear perturbation to the metric $g_{ab}$, which is a solution of Eq. \eqref{scg}. Note that $h_{ab}$ is implicitly assumed to be a (classical) tensor stochastic field. One could also add higher order corrections to the background geometry by taking into account higher order stress-energy fluctuations.

An important point regarding the SSCG approach is that, in order to ensure the consistency of Eq. \eqref{sscgeq}, $\xi_{ab}$ must satisfy $\nabla^a \xi_{ab} = 0$ (with $\nabla^a$ the covariant derivative associated with the background metric $g_{ab}$). In Ref. \cite{verdaguer1998b} it is shown that the fact that $\nabla^a \hat T_{ab} = 0$ implies that $\nabla^a_x N_{abcd} (x,y) = 0$. Therefore, applying the covariant derivative to the correlation functions in Eq. \eqref{sscgeq}, one gets $\bra \nabla^a \xi_{ab} \ket_s = 0 $ and $\bra \nabla^a_x \xi_{ab} (x) \nabla^c_y \xi_{cd} (y) \ket_s = 0 $. From this, according to Refs. \cite{verdaguer1998b,verdaguer1999a}, one concludes that $\nabla^a \xi_{ab}$ is deterministic and equal to the zero vector field, guaranteeing the consistency of SSCG.

The problem is that the previous argument is not solid. Consider for simplicity a classical stochastic scalar variable $\varphi(x)$ that the can only take the values +1 or -1. Let us further assume that we have a distribution such that, at each point, $\varphi$ takes those two values with equal probability, and without correlations between the values at two distinct points. Clearly, after $N \to \infty$ realizations of $\varphi(x)$, the statistical average of the 1-point function vanishes, i.e., $\bra \varphi(x) \ket_s =0$. Let us now focus on the product $\varphi(x)\varphi(y)$. The only two possible values for such a product are either -1 or +1, and both occur with equal probability. Hence, after $N \to \infty$ realizations, the statistical average of the 2-point function also vanishes, i.e., $\bra \varphi(x)\varphi(y) \ket_s = 0 $. Thus, in analogy with Eq. \eqref{sscgeq}, we have that $\bra \varphi(x) \ket_s =0$ and $\bra \varphi(x)\varphi(y) \ket_s=0$; nevertheless, $\varphi(x)$ is a never-vanishing stochastic field, and it is completely non-deterministic. This is a clear counter example for the argument in the latter paragraph. Namely, the fact that $\bra \nabla^a \xi_{ab} \ket_s = 0 $ and $\bra \nabla^a_x \xi_{ab} (x) \nabla^c_y \xi_{cd} (y) \ket_s = 0 $, does not necessarily imply that $\nabla^a \xi_{ab}$ is deterministic and equal to the zero vector field.\footnote{Note that the missing aspect of the analysis would be the consideration of $\varphi(x)\varphi(x)$ and its statistical average. If one could argue that such a quantity vanishes, one would have grounds to argue that $\varphi(x)$ might be deterministic and equal to zero. Of course, that does not hold in our example, where $\varphi(x)\varphi(x)=1$ on every element of the ensemble, and so its average value. Regarding the situation at hand, one would have to find a way to argue that $\bra \nabla^a_x \xi_{ab} (x) \nabla^c_x \xi_{cd} (x) \ket_s = 0 $ and there seems to be no path for doing so.} As a result, given that $\nabla^a \xi_{ab} =0 $ does not necessarily hold for every realization of $\xi_{ab}$, is clear that Eq. \eqref{sscg} cannot be valid for every $\xi_{ab}$.
This in turn implies that Eq. (\ref{sscg}) is inconsistent for the generic individual realizations, thus undercutting the program as a whole.

The SSCG considered in the WZU model is not the same as the one characterized by Eqs. \eqref{sscg} and \eqref{sscgeq}, i.e. the WZU's model is not based on conventional SSCG.  One particular difference between the two approaches is that the SSCG framework characterized by Eq. \eqref{sscg} relies on a perturbative analysis. The WZU model, on the other hand, deals with a situation that is extremely inhomogeneous (with $(\bra \hat T^2_{00} \ket - \bra \hat T_{00} \ket^2 ) \sim \bra \hat T_{00} \ket^2 \sim \Lambda^8$), implying that its gravitational effects cannot be treated perturbatively. Recall that the approach used in the WZU model employs Einstein's equations with the matter fields regarded as classical stochastic fields, with their statistical properties determined by quantum expectations values. A such, the approach can be regarded as a nonperturbative version of the SSCG scheme described above.

However, conventional SSCG and WZU's version of SSCG share the same difficulty, namely that applying the 4-divergence to each side of the Einstein field equations (EFE) yields an inconsistency. In WZU's model, the corresponding EFE $G_{ab} = 8 \pi G T_{ab}$ with metric \eqref{frw2} lead to an inhomogeneous type of Friedmann equations. On the other hand the scale factor $a(\x,t)$ and the stress-energy tensor $T_{ab}$ are interpreted as classical stochastic fields. In particular, the statistical properties of $T_{ab}$ are supposed  to be inherited from the quantum expectation values, e.g. $\bra \hat T_{ab} \ket$ and Eq. \eqref{covTab}. For example, the statistical mean of the 00 component of stress-energy tensor, $\bra T_{00} (x) \ket_s$, is related to $\bra \hat T_{00} (x) \ket \simeq \Lambda^4$. Therefore, given that $\nabla^a \bra T_{ab} (x) \ket_s=0$, the 4-divergence of the statistical mean of $T_{ab}$ vanishes. At this point, the same problem that arises in conventional SSCG appears in WZU's case. That is, what enters into WZU's version of EFE is not the average of the stochastic field, in this case $\bra T_{ab} \ket_s$, but a realization of a particular $T_{ab} (x)$ at each point $x$. Consequently, quite generically we would have $\nabla^a T_{ab} (x) \neq 0$, but for any metric and in particular that of \eqref{frw2} we have $\nabla^a G_{ab} (x) = 0$; hence one has a deeply problematic result, because it implies that the semiclassical equations considered are simply inconsistent in the context of the premises of the setting at hand. Note also that in WZU's model one cannot claim that what appears in the right hand side of EFE is $\bra \hat T_{ab} \ket \simeq \Lambda^4$ since that is a homogeneous quantity. On the other hand, in WZU's model the left hand side of EFE is supposed to be associated in a manner not clearly specified with quantum expectation values of the metric degrees of freedom, and in fact, that leads to ambiguities in their final results (near the end of the next subsection we will be more specific about that latter issue).

We end this subsection by pointing out that the aforementioned analysis is independent of the discussion presented in Sec. \ref{secQF}. That is, even if one were to somehow  accept that ``vacuum quantum fluctuations'' generate an inhomogeneous stochastic $T_{ab}$, there are inconsistencies in the WZU's version of SSCG when using such stress-energy tensor as a matter source in EFE.


\subsection{Some issues with the actual calculations underlying the WZU proposal}

To conclude our critique of the WZU proposal, let us ignore for the meantime the previous objections and grant i) that large quantum fluctuations imply a highly inhomogeneous vacuum energy density and ii) that such a density, modeled as a classical stochastic field, can be adequately employed as source in the Einstein equations. In what follows, we will show the actual calculations performed by WZU to be incompatible with these granted assumptions.

The first point we would like to make is that if, as argued by WZU, the stress-energy variation from a spacetime point to its neighbors is generically as large as the stress-energy at the point itself, then it is unreasonable to constrain the spacetime metric to have the particular simple form of Eq. \eqref{frw2}. Such a metric has only one degree of freedom per spacetime point, rather than the 6 generic degrees of freedom of an unconstrained metric, so it seems incorrect to expect it to satisfy Einstein's equations associated with a random stress-energy tensor. In fact, even if that was the form of the metric at some initial time (i.e., if one is given initial data compatible with such a form at some initial hypersurface), the extremely large variations of the stress-energy tensor to the future of that hypersurface would rapidly modify the spacetime metric producing large inhomogeneous terms. A concrete problem emanating from all this is the following.

In the WZU model, the solution of Eq. \eqref{masterunruh} is of the form $a(\x,t)= e^{H t} P(\x,t)$, with $P$ a quasi-periodic function in $t$. Let us now focus on the initial constraints given by Einstein's equations with components $0i$ ($i=1,2,3$). Using the solution for $a$, these equations yield
\begin{equation}\label{EE0i}
\partial_i \left( \frac{\dot P}{ P } \right) = -4\pi G J_i,
\end{equation}
where we have introduced the notation $\textbf{J} \equiv (T_{01},T_{02},T_{03}) $. Now, applying $\varepsilon^{ijk} \partial_j $ to both sides of the equation, with $\varepsilon^{ijk}$ the Levi-Civita tensor, leads to a problem: while the left-hand side vanishes automatically, there is no reason for the right-hand side to do so. By assumption, we have a highly inhomogeneous and anisotropic $T_{ab}$, fluctuating randomly from point to point, so there is no reason for it to satisfy $\varepsilon^{ijk} \partial_j T_{0k}=0$.
Thus, there is an incompatibility between the following assumptions: A) a highly inhomogeneous and fluctuating stress-energy tensor, and B) the metric ansatz \eqref{frw2}.

The authors of the WZU model agree that the use of the simple inhomogeneous metric of Eq. \eqref{frw2} might result in inconsistencies. They argue, however, that one should take the results obtained with it as a ``first approximation''. In principle, we agree with the spirit of such a proposal (and recalling that for the present analysis we have left out the problems presented in Sects.  \ref{secQF} and \ref{secSSCG}), i.e. that a non-pertubative computation in the metric and the stress-energy operator may not be feasible for practical reasons. However, the approximation scheme one is supposed to be applying to the problem at hand should, at least, involve a clearly identified small expansion parameter as well as some well defined scheme allowing for instance the possibly of studying the backreaction effects in each step of the approximation. The problem in WZU's model is that in practice there is none of that in their supposed approximation. The model deems $T_{ab}$ to be extremely inhomogeneous and anisotropic. Hence, in some sense, the stress-energy tensor is considered in a completely non-perturbative way in WZU's model when the intent is just the opposite.  Moreover, there is no small parameter present that could characterize the alleged approximation. Therefore, the predictions extracted from this formalism cannot really be considered approximations to any well-defined quantities, so they cannot be trusted. Specifically, we see no reason to trust the results derived from WZU main equation, i.e. Eq. \eqref{masterunruh}, but at the same time dismiss the inconsistency derived from Eq. \eqref{EE0i}, provided that both equations are obtained from the same version of EFE.

In \cite{unruh2017}, more general metrics are analyzed and new arguments are presented to support the estimates obtained from the metric of Eq. \eqref{frw2}. The problem, as we explain bellow, is that those arguments rely on dubious identifications between observable quantities, quantum expectation values and ensemble averages.

For instance, in \cite{unruh2017} one finds
\begin{equation}
\frac{\dot a}{a } (\x,t) = \frac{\dot a}{a } (\x_0,t) - 4 \pi G \int_{\x_0}^{\x} \textbf{J} (\x',t) \cdot \textbf{dl}'
\end{equation}
where $\textbf{dl}' = (dx',dy',dz')$ and $\x_0$ is an arbitrary spatial point. According to the WZU treatment,
from the above equation, since $\bra \textbf{J} \ket =0$, one concludes that $\bra \dot a/a \ket_s = 0$. Such a result is then interpreted as cancellations of local contractions and expansions of the spacetime sourced by ``quantum fluctuations''. Another example is the estimation of the difference between the values of $\dot a /a$ at two fixed spatial points separated by a distance of order $\Delta x \sim 1/\sqrt{G} \Lambda^2$. In \cite{unruh2017} one finds
\begin{equation}\label{EE0i2}
\Delta \left( \frac{\dot a}{a } \right) \sim 4\pi G J \Delta x \sim \sqrt{G} \Lambda^2 \sim \sqrt{\left\bra \left( \frac{\dot a}{a } \right)^2 \right\ket_s } ,
\end{equation}
in which the first estimation is essentially Eq. \eqref{EE0i}. Now, the second estimation comes from associating the classical value of $J$ with the square root of its quantum fluctuation, i.e., $J = \sqrt{\bra \hat J^2 \ket} \sim \Lambda^4$, and the last estimation is given by associating the square root of the quantum expectation value $\sqrt{ \bra \hat T_{00} \ket } \sim \Lambda^2$ with the square root of the classical ensemble average of the quantity $(\dot a/a)^2$ (this last association is made via Einstein's equations with components $00$, which is explicitly given by $ \bra G_{00} \ket_s = 8 \pi G \bra \hat T_{00} \ket$).

The point we would like to make is that estimations of physical observables from quantum expectation values, such as those described above, most be handled with care---particularly in the cosmological context at hand. It is clear that quantum expectation values cannot be directly associated with measurement results, only with averages of measurement results performed over ensembles of identically prepared systems. In a cosmological setting, such ensembles are nowhere to be found. Moreover, it is often the case that the expectation value is not even a possible value for the result of a measurement, which always has to be an eigenstate of the measured quantity.

To conclude we note that, as we have mentioned, the SSCG used by WZU differs from traditional SCG or the SSCG represented in Eq. \eqref{sscg}. In particular, in the WZU model, the gravitational degrees of freedom are treated as quantum operators in some parts of the calculation. This is illustrated, for example, by the computation of the Wigner transform of $\widehat{\dot a/a}$ in order to obtain the quantum expectation value $\bra \widehat{\dot a/a} \ket$, as shown in Eq. \eqref{wtdotaa}.\footnote{ Note that the use of Wigner's quasi-distribution function automatically implies that one is going to compute an expectation value of a quantum operator. In other words, Wigner's function generically possess negative values, therefore it cannot be taken to represent in any sense a probability distribution function for a classical variable.} It is worth noting, as mentioned in \cite{unruh2018}, that the results would change if, instead of considering $\bra \widehat{\dot a/a} \ket$, one focuses on $\bra \dot{\hat{a}} \ket / \bra \hat a \ket$. These ambiguities disappear when adopting a SCG framework in which gravity is always classical but the matter fields are subjected to a QFT description.

\section{Conclusions}
\label{conclusions}
Naive vacuum energy estimates of the value of the cosmological constant produce results that are several orders of magnitude larger than those extracted from cosmological observations. Severe  fine tuning of the bare cosmological constant  seems to  be required to deal with that.  Obtaining a deeper understanding of this puzzle (and perhaps to achieve a final solution) is one of the major challenges of modern physics. This is because the cosmological constant problem  is  likely connected with several aspects of theoretical physics that can still be considered open issues. In particular, the resolution of the cosmological constant problem might be related with: a complete theory for interacting quantum fields and  renormalization  in a curved space-time (see for instance \cite{Shapiro}); a full workable theory of quantum gravity, and/or perhaps  other topics  \cite{Josset16,Perez17}.

The WZU proposal is a valiant attempt to deal with the cosmological constant problem, one of the most challenging issues at the interface between gravity and quantum theory. Unfortunately, the proposal is beset by several devastating problems. The difficulties involve issues that touch on the conceptual framework of quantum theory, its application to the cosmological setting, various self-consistency concerns within stochastic semiclassical gravity and problematic aspects in the actual calculations.

Recapitulating in some detail, we have argued that:
\begin{enumerate}
\item[I)] The claim that the quantum fluctuations of the vacuum imply a highly inhomogeneous vacuum energy density, which is a central tenet of the WZU construction, is simply inconsistent with the standard interpretation of quantum mechanics.
\item[II)] The implementation of stochastic semiclassical gravity is not self-consistent because there is no mechanism at play to ensure that the stress-energy tensor satisfies the conservation equation in each realization of the stochastic process. Therefore, the stress-energy tensor is incompatible as a source in Einstein's field equations.
\item[III)] The equation employed to describe spacetime within the WZU model is in fact inconsistent with the postulated, highly fluctuating nature of the stress-energy tensor. Henceforth, there is no reason to accept the results derived from one of the equations (which claim to obtain the correct magnitude of the cosmological constant), while at the same time, one must clearly recognize the inconsistency of another one of the main set of equations employed.
\end{enumerate}

We stress that items II and III hold independently of item I (which some people may discard as ``just philosophy'' or plainly reject it based on their preconceptions regarding the foundations on Quantum Mechanics). That is, even if one were to accept all the premises in WZU's model (see Sec. \ref{unruhmodel}), items II and III reveal inconsistencies in such a model. It is thus the inescapable conclusion that the WZU proposal, at least in its present form, is in fact inadequate as a solution to the cosmological constant problem.
\vspace*{1cm}
\begin{acknowledgements}
G.R.B. is supported by CONICET (Argentina) and he acknowledges support from grant PIP 112-2017-0100220CO of CONICET (Argentina). G.L. is supported by CONICET (Argentina) and the National Agency for the Promotion of Science and Technology (ANPCYT) of Argentina grant PICT-2016-0081. D.S. acknowledges support during the elaboration of this manuscript from the FAE-Network of CONACYT, as well as the sabbatical fellowships from CO-MEX-US (Fullbright-Garcia Robles) and from DGAPA-UNAM (Paspa). E.O. acknowledges support from UNAM-PAPIIT grant IN102219.
\end{acknowledgements}
\bibliography{bibliografia}
\bibliographystyle{apsrev}
\end{document}